\documentclass[11pt]{article}
\usepackage{graphicx}

\newcommand{\BABARPubYear}    {00}

\newcommand{\BABARProcNumber} {37}
\newcommand{\SLACPubNumber} {8727}

\def\lbabar{\mbox{{\large\sl B}\hspace{-0.4em} {\normalsize\sl A}\hspace{-0.03em}{\large\sl B}\hspace{-0.4em} {\normalsize\sl A\hspace{-0.02em}R}}}
\usepackage{relsize}
\def\babar{\mbox{\slshape B\kern-0.1em{\smaller A}\kern-0.1em
    B\kern-0.1em{\smaller A\kern-0.2em R}}}

\def\epem       {\ensuremath{e^+e^-}}

\def\pip   {\ensuremath{\pi^+}}

\def\Kbar  {\kern 0.2em\overline{\kern -0.2em K}{}}

\def\KS    {\ensuremath{K^0_{\scriptscriptstyle S}}} 
 
\def\Kstarz  {\ensuremath{K^{*0}}}

\def\Kzb   {\ensuremath{\Kbar^0}}
\def\KzKzb {\ensuremath{K^0 \kern -0.16em \Kzb}}

\def\Dbar  {\kern 0.2em\overline{\kern -0.2em D}{}}

\def\Dzb   {\ensuremath{\Dbar^0}}
\def\DzDzb {\ensuremath{D^0 {\kern -0.16em \Dzb}}}

\def\Bz    {\ensuremath{B^0}}
\def\B     {\ensuremath{B}}
\def\Bbar  {\kern 0.18em\overline{\kern -0.18em B}{}}

\def\Bzb   {\ensuremath{\Bbar^0}}
\def\Bu    {\ensuremath{B^+}}
\def\Bub   {\ensuremath{B^-}}

\def\BzBzb {\ensuremath{B^0 {\kern -0.16em \Bzb}}}

\def\jpsi  {\ensuremath{{J\mskip -3mu/\mskip -2mu\psi\mskip 2mu}}} 
\def\psitwos {\ensuremath{\psi{(2S)}}}
\mathchardef\Upsilon="7107
\def\Y#1S{\ensuremath{\Upsilon{(#1S)}}}

\def\FourS {\Y4S}

\mathchardef\Deltares="7101
\mathchardef\Xi="7104
\mathchardef\Lambda="7103
\mathchardef\Sigma="7106
\mathchardef\Omega="710A
\def\Deltabar   {\kern 0.25em\overline{\kern -0.25em \Deltares}{}}
\def\Lbar {\kern 0.2em\overline{\kern -0.2em\Lambda\kern 0.05em}\kern-0.05em{}}
\def\Sigbar{\kern 0.2em\overline{\kern -0.2em \Sigma}{}}
\def\Xibar{\kern 0.2em\overline{\kern -0.2em \Xi}{}}
\def\Obar{\kern 0.2em\overline{\kern -0.2em \Omega}{}}
\def\Nbar{\kern 0.2em\overline{\kern -0.2em N}{}}
\def\Xbar{\kern 0.2em\overline{\kern -0.2em X}{}}

\def\mes        {\mbox{$m_{\rm ES}$}}

\def\ev   {\ensuremath{\rm \,e\kern -0.08em V}}
\def\kev  {\ensuremath{\rm \,ke\kern -0.08em V}} 
\def\mev  {\ensuremath{\rm \,Me\kern -0.08em V}} 
\def\gev  {\ensuremath{\rm \,Ge\kern -0.08em V}} 
\def\gevc {\ensuremath{{\rm \,Ge\kern -0.08em V\!/}c}} 
\def\tev  {\ensuremath{\rm \,Te\kern -0.08em V}}
\def\mevc {\ensuremath{{\rm \,Me\kern -0.08em V\!/}c}} 
\def\gevcc{\ensuremath{{\rm \,Ge\kern -0.08em V\!/}c^2}} 
\def\mevcc{\ensuremath{{\rm \,Me\kern -0.08em V\!/}c^2}}

\def\cm   {\ensuremath{\rm \,cm}}

\def\mum  {\ensuremath{\,\mu\rm m}} 

\def\invpb {\ensuremath{\mbox{\,pb}^{-1}}}

\def\invfb   {\ensuremath{\mbox{\,fb}^{-1}}}
\def\mus  {\ensuremath{\rm \,\mus}}

\def\ps   {\ensuremath{\rm \,ps}}

\def\mus        {\ensuremath{\,\mu{\rm s}}}    
\def\ps         {\ensuremath{{\rm \,ps}}}   
\def\gsim{{~\raise.15em\hbox{$>$}\kern-.85em
          \lower.35em\hbox{$\sim$}~}}
\def\lsim{{~\raise.15em\hbox{$<$}\kern-.85em
          \lower.35em\hbox{$\sim$}~}}

\def\CP                 {\ensuremath{C\!P}}

\def\to                 {\ensuremath{\rightarrow}}

\def\pep2{PEP-II}
\def\BF{$B$ Factory}

\def\sb{${\sin\! 2 \beta   }$}

\def\stwob{\ensuremath{\sin\! 2 \beta   }}

\def\mistag{\ensuremath{w}}

\def\deltaz{\ensuremath{{\rm \Delta}z}}
\def\deltat{\ensuremath{{\rm \Delta}t}}
\def\deltamd{\ensuremath{{\rm \Delta}m_d}}

\providecommand{\eqref}[1]{Eq.~(\ref{eq:#1})}

\newcommand{\epjc}      [1]  {{Eur.\ Phys.\ Jour.\ C~{\bf #1}}}

\def\jetset74   {\mbox{\tt Jetset \hspace{-0.5em}7.\hspace{-0.2em}4}}

\def\result {\ensuremath{\stwob=0.12\pm 0.37\, {\rm (stat)} \pm 0.09\, {\rm (syst)}}}
\providecommand{\btodstarlnu}{\mbox{$B\to D^{*}\ell\nu$}}
\providecommand{\bztodstarlnu}{\mbox{$B^0\to D^{*-}\ell^+\nu$}}

\long\def\inst#1{\par\nobreak\kern 4pt\nobreak
    {\it #1}\par\vskip 10pt plus 3pt minus 3pt}

\begin{document}
{\pagestyle{empty}

\begin{flushright}
SLAC-PUB-\SLACPubNumber \\
\babar-PROC-\BABARPubYear/\BABARProcNumber \\
December, 2000 \\
\end{flushright}

\par\vskip 4cm

\begin{center}
\Large \bf First Physics Results at {\mbox{\slshape B\kern-0.1em{\smaller A}\kern-0.1em B\kern-0.1em{\smaller A\kern-0.2em R}}}
\end{center}
\bigskip

\begin{center}
\large 
Francesca Di Lodovico       \\
University of Edinburgh  \\
Dept. of Physics \\
James Clerk Maxwell Bldg., King's Bldgs \\
Mayfield Road \\
Edinburgh EH9 3JZ, Scotland, UK \\
(for the \lbabar\ Collaboration)
\end{center}
\bigskip \bigskip

\large 
The \babar\ detector, which operates at the SLAC \pep2\ asymmetric \epem\ 
collider at energies near the \FourS\ resonance, started operation
on the 26$\rm ^{th}$ of May 1999. We present the first study of \sb, with samples 
of $\Bz \to \jpsi \KS$ and $\Bz \to \psitwos \KS$ decays, using 
$9.0 \invfb$ of data recorded between January and July 2000 at the \FourS\ 
resonance and $0.8 \invfb$ recorded 40\mev\ below the \FourS\ resonance. 
A preliminary result of \result\  was obtained. 
Details of the analysis are given. 
Moreover, we present measurements of charged and neutral \B\ meson lifetimes and $\Bz\Bzb$  
oscillation frequency.

\vfill
\begin{center}
Contributed to the Proceedings of the 5$^{th}$ Heavy Quarks at Fixed Target Conference, \\
10/09/2000---10/12/2000, Rio de Janeiro, Brazil
\end{center}

\vspace{1.0cm}
\begin{center}
{\em Stanford Linear Accelerator Center, Stanford University, 
Stanford, CA 94309} \\ \vspace{0.1cm}\hrule\vspace{0.1cm}
Work supported in part by Department of Energy contract DE-AC03-76SF00515.
\end{center}

\section{Introduction}
The primary goal of the \babar\ experiment at \pep2\ is to overconstrain the Unitarity Triangle.
The sides of this triangle can be measured through non-\CP\ violating physics, such as  
$V_{ub}$, $V_{cb}$, $V_{td}$ measurements~\cite{PhysBook}, 
while its angles are accessible through \CP\ violating processes~\cite{PhysBook}.
\section{PEP-II}
The \pep2\ $B$ Factory~\cite{BabarPub0018} is an \epem\ colliding beam storage ring complex on the SLAC site 
designed to produce a luminosity of at least 3x$10^{33} \cm^{-2}s^{-1}$ at 
a center--of--mass energy of 10.58\gev, the mass of the \FourS\ resonance. In the 2000 run, 
the achieved average luminosity was 2.5x$10^{33} \cm^{-2}s^{-1}$, with a daily 
average integrated luminosity of $135 \invpb$. The total collected luminosity
was about $22 \invfb$.
The machine is asymmetric with a High Energy Ring (HER) for the 9.0\gev\ electron beam
and a Low Energy Ring (LER) for the 3.1\gev\ positron beam. This corresponds to
$\rm {\beta\gamma}$=0.56 and makes it possible to measure time dependent \CP\ violating asymmetries. 
It corresponds to an average separation of $\rm {\beta\gamma c \tau}$=250\mum\ 
between the two B mesons vertices.
\section{\mbox{\sl B\hspace{-0.4em} {\small\sl A}\hspace{-0.4em} \sl B\hspace{-0.4em} {\small\sl A\hspace{-0.1em}R}}}
\subsection{Detector description~\cite{BabarPub0018}}
The volume within the \babar\ superconducting solenoid, which produces a 1.5 T axial magnetic
field, consists of: a five layer silicon strip vertex detector (SVT), a central drift
chamber (DCH), a quartz-bar Cherenkov radiation detector (DIRC) and a CsI
crystal electromagnetic calorimeter (EMC). Two layers of
cylindrical resistive plate counters (RPCs) are located between the barrel
calorimeter and the magnet cryostat. All the detectors located inside the
magnet have full acceptance in azimuth. The integrated flux return (IFR) outside the cryostat is
composed of 18 layers of steel, which successively increase in thickness away from the
interaction point, and are instrumented with 19 layers of planar RPCs in the barrel and 18 in the endcaps.
\subsection{Event reconstruction~\cite{BabarPub0018}}
Charged particles are detected and their momentum is measured by a combination of 
the DCH and SVT. The  charged particle momentum resolution is approximately given by 
$\left( \delta p_T / p_T \right)^2 =  (0.0015\, p_T)^2 + (0.005)^2$, where 
$p_T$ is in \gevc.  The SVT, with a typical resolution of 
10\mum\ per hit, provides excellent vertex resolution  both in 
the transverse plane and in $z$.  The vertex resolution in $z$ is typically 50~\mum\ for 
a fully reconstructed \B\ meson and of order 100~\mum\ for the distance 
among the two \B\ mesons when only one is fully reconstructed.
Leptons and hadrons are identified using a combination of measurements
from all the \babar\ components, including 
the energy loss ${\rm d}E/{\rm d}x$ in the helium-based 
gas of the DCH (40 samples maximum) and in the silicon of the SVT (5 samples
maximum). 
Electrons and photons are identified in the barrel and the forward regions 
by the EMC, and muons are identified in the IFR. In the barrel region the DIRC
provides excellent kaon identification over the full 
momentum range above 250\mev/c.
\section{${\sin\! 2 \beta   }$ measurement}
In \epem\ storage rings operating at  the \FourS\ resonance a \BzBzb\ pair
produced in a \FourS\ decay 
evolves in a coherent $P$-wave until one of the \B\ mesons decays. 
If one of the \B\ mesons ($B_{tag}$) can be 
ascertained to decay to a state of known flavor at a certain time $t_{tag}$, 
the other \B\  ($B_{CP}$) is {\it at that time} known to be of the opposite flavor.
For the measurement of \sb, $B_{CP}$ is fully reconstructed in a \CP\ 
eigenstate ($\jpsi \KS$ or $\psitwos \KS$).
By measuring the proper time interval $\deltat = t_{CP} - t_{tag}$ from the $B_{tag}$ decay time 
to the decay of the $B_{CP}$ ($t_{CP}$), it is possible to determine the time evolution 
of the initially pure \Bz\ or \Bzb\ state:
\begin{equation}
\label{eq:TimeDep}
        f_\pm(\, \deltat \, ; \,  \Gamma, \, \deltamd, \, {\cal {D}} \sin{ 2 \beta } )  = {\frac{1}{4}}\, \Gamma \, {\rm e}^{ - \Gamma \left| \deltat \right| }\, \left[  \, 1 \, \pm \, {\cal {D}} \sin{ 2 \beta } \times \sin{ \deltamd \, \deltat } \,  \right]\ ,
\end{equation}
where the $+$ or $-$  sign 
indicates whether the 
$B_{tag}$ is tagged as a \Bz\ or a \Bzb, respectively.  The dilution factor ${\cal {D}}$ is given by 
$ {\cal {D} } = 1 - 2 \mistag$, where $\mistag$ is the mistag fraction, {\it i.e.}, the 
probability that the flavor of the tagging \B\ is identified incorrectly. 
A direct \CP\ violation term proportional to $\cos {  \deltamd \, \deltat }$ 
could arise from the interference between two decay mechanisms with different
weak phases.  In the Standard Model, we consider that the dominant diagrams for the decay  modes 
have no relative weak phase, so no such term is expected.
\par
To account for the finite resolution of the detector,
the time-dependent distributions $f_\pm$ for \Bz\ and \Bzb\ tagged events 
(Eq.~\ref{eq:TimeDep}) must be convoluted with 
a time resolution function ${\cal {R}}( \deltat ; \hat {a} )$:
\begin{equation}
\label{eq:Convol}
        {\cal F}_\pm(\, \deltat \, ; \, \Gamma, \, \deltamd, \, {\cal {D}} \sin{ 2 \beta }, \hat {a} \, )  = 
f_\pm( \, \deltat \, ; \, \Gamma, \, \deltamd, \, {\cal {D}} \sin{ 2 \beta } \, ) \otimes 
{\cal {R}}( \, \deltat \, ; \, \hat {a} \, ) \ ,
\end{equation}
where $\hat {a}$ represents the set of parameters that describe the resolution function.  
\par
It is possible to construct a \CP-violating observable 
\begin{equation}
  {\cal A}_{CP}(\deltat) = \frac{ {\cal F}_+(\deltat) \, - \, {\cal F}_-(\deltat) }
{ {\cal F}_+(\deltat) \, + \, {\cal F}_-(\deltat) } \ \ , 
\label{eq:asymmetry}
\end{equation}
which is approximately proportional to  \stwob:
\begin{equation}
 {\cal A}_{CP}(\deltat) \sim {\cal D} \sin{ 2 \beta } \times \sin{ \deltamd \, \deltat } \ \ .
\label{eq:asymmetry2}
\end{equation}
Since no time-integrated \CP\ asymmetry effect is expected, an
analysis of the time-dependent asymmetry is necessary.  
At an asymmetric-energy \BF, the proper decay-time difference $\deltat$ is, 
to an excellent approximation, proportional to 
the distance \deltaz\ between  the two \Bz-decay vertices 
along the axis of the boost, 
$\deltat \approx \deltaz / {\rm c} \left< \beta \gamma \right>  $.  
\par
Since the amplitude of the time-dependent \CP -violating 
asymmetry in Eq.~\ref{eq:asymmetry2} 
involves the product of ${\cal {D}}$ and \stwob, one needs to determine 
the dilution factors ${\cal {D}}_i $ (or equivalently the mistag fractions $\mistag_i$) 
in order to extract the value of \stwob.
The mistag fractions are determined from the data
by studying the time-dependent rate of \BzBzb\ oscillations.
\par
The value of the single free parameter \stwob\ is extracted from the 
tagged $B_{CP}$ sample
by maximizing the likelihood function
\begin{equation}
\label{eq:Likelihood}
 \ln { {\cal {L} }_{CP} } = \sum_{i}
\left[ \, \sum_{\Bz \, {\rm tag} } {  \ln{ {\cal F}_+( \, \deltat \, ; \,  \Gamma, \, \deltamd, \, \hat {a}, 
\, {\cal {D}}_i \sin{ 2 \beta }  \,) } } +
\sum_{\Bzb \, {\rm tag} }{ \ln{ {\cal F}_-( \, \deltat \, ; \,  \Gamma, \, \deltamd, \,  \hat {a}, \, {\cal {D}}_i \sin{ 2 \beta }  \, ) } } \, \right] \ ,
\end{equation} 
where the outer summation is over tagging categories $i$.

\subsection{Analysis}
For this analysis we use a sample of $9.8 \invfb$ of data recorded
between January and the beginning of July 2000, 
of which $0.8 \invfb$ was recorded 40\mev\ below the \FourS\ resonance (off-resonance data). 
\par
The measurement of the \CP-violating asymmetry has five main components~:
\begin{itemize}
\item
Selection of the signal $\Bz/\Bzb \to \jpsi \KS$ 
and $\Bz/\Bzb \to \psitwos \KS$ events, as described in detail
in~\cite{BabarPub0005}.   
\par
Distributions of ${\rm \Delta} E$, the difference between the reconstructed and
expected \B\ meson energy measured in the center--of--mass frame, and \mes,
the beam--energy substitued mass, are shown 
in Fig.~\ref{fig:jks00} for the $\jpsi \KS$ ($\KS \to \pi^+ \pi^-$) 
and $\jpsi \KS$ ($\KS \to \pi^0 \pi^0$) samples. 
Signal event yields and purities, determined 
from a fit to the \mes\ distributions after selection 
on ${\rm \Delta} E$, are summarized in
Table~\ref{tab:CharmoniumYield}. 

\begin{figure}[t]
    \centering
 \begin{tabular}{lr}   
    \includegraphics[height=8cm,width=8cm]{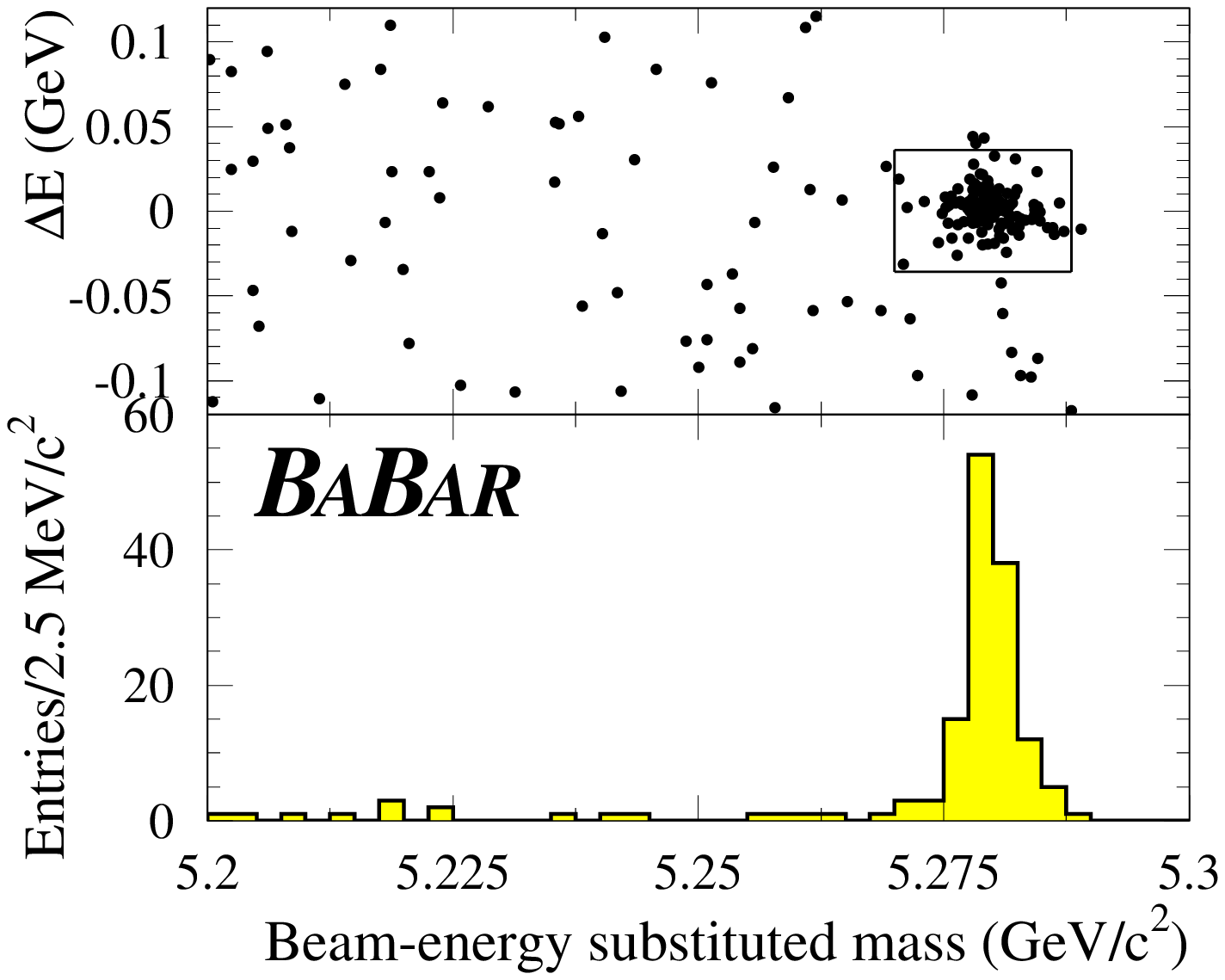}
    \includegraphics[height=8cm,width=8cm]{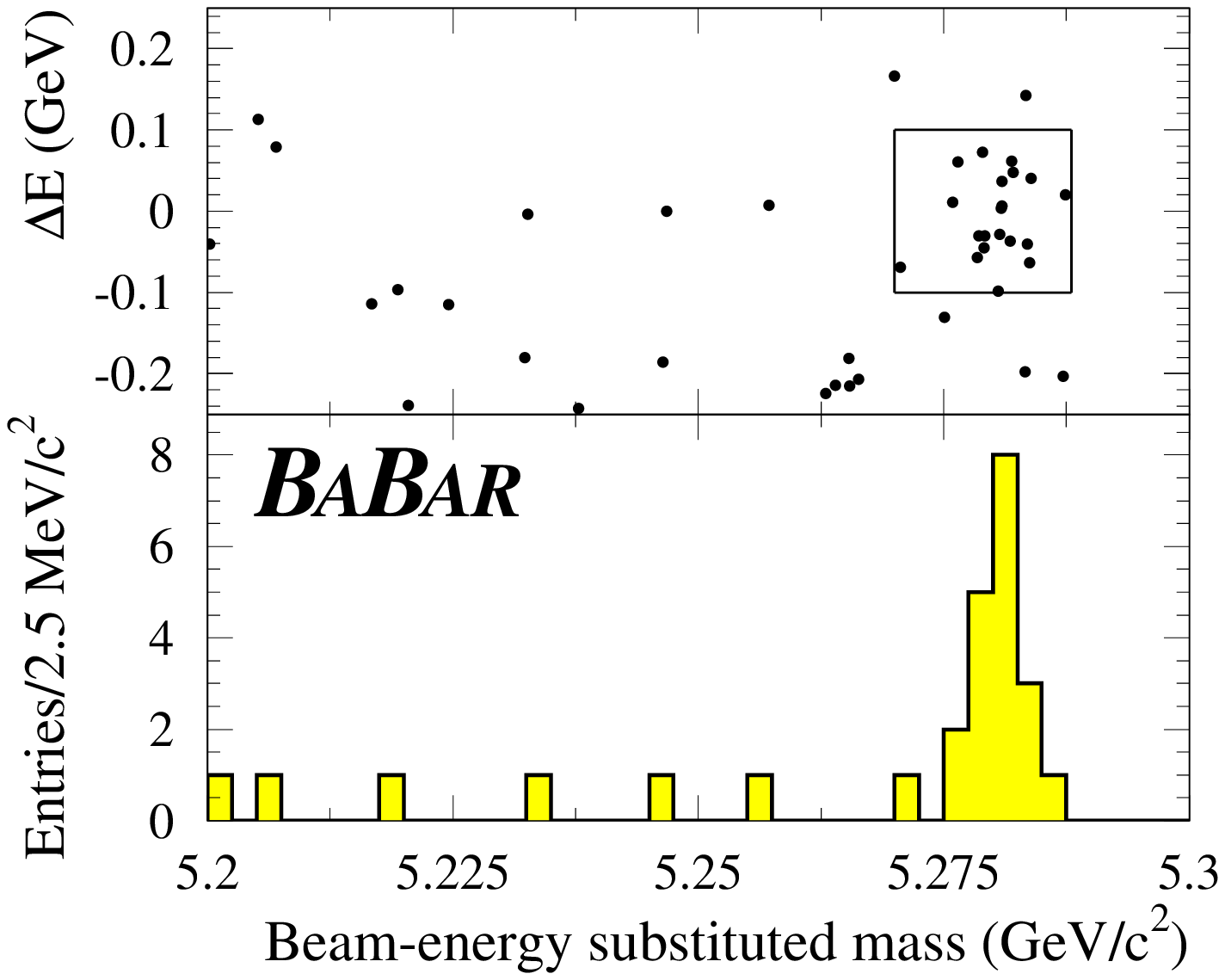}
\end{tabular}
 \caption{\it $\jpsi \KS$ ($\KS \to \pi^+ \pi^-$) signal (left).
$\jpsi \KS$ ($\KS \to \pi^0 \pi^0$) signal (right).}
    \label{fig:jks00} 
\end{figure}\medskip

\begin{table}[!htb]
\caption{
Event yields for the different samples used in this analysis, 
from the fit to \mes\ distributions after selection 
on ${\rm \Delta} E$.  The purity is quoted for 
$\mes >5.270 \mevcc$. 
} 
\vspace{0.3cm}
\begin{center}
\begin{tabular}{|l|c|c|} \hline
Final state  & Yield & Purity (\%) \\ \hline \hline
$\jpsi \KS$ ($\KS \to \pi^+\pi^-$)              &  124$\pm$12&  96 \\
$\jpsi \KS$ ($\KS \to \pi^0 \pi^0$)             &   18$\pm$4& 91 \\
$\psitwos \KS$                                  &   27$\pm$6&  93 \\ 
\hline
\end{tabular}
\end{center}
\label{tab:CharmoniumYield}
\end{table}

\item
Measurement of the distance \deltaz\ between the two \Bz\ decay 
vertices along the \FourS\ boost axis, as described in detail in~\cite{BabarPub0008} and~\cite{BabarPub0007}.
\par
From the measurement of \deltaz\ , \deltat\ can be computed.
\par
The time resolution function is described accurately by the sum of
two Gaussian distributions, which has five independent parameters:
\begin{eqnarray}
{\cal {R}}( \, \deltat ; \, \hat {a} \,  ) &=&  \sum_{i=1}^{2} { \, \frac{f_i}{\sigma_i\sqrt{2\pi}} \, {\rm exp} \left(  - ( \deltat-\delta_i )^2/2{\sigma_i }^2   \right) } \, \, .
\end{eqnarray}
A fit to the time resolution function in Monte Carlo simulated events indicates 
that most of the events ($f_1 = 1-f_2 = 70\%$) are in the core 
Gaussian, which has a width $\sigma_1 \approx 0.6 \ps$.
The wide Gaussian has a width  $\sigma_2 \approx 1.8\ps$. 
Tracks from forward-going charm decays included in the reconstruction of the $B_{tag}$ 
vertex introduce a small bias, $\delta_1 \approx -0.2 \ps$, for the core Gaussian.  
\item
Determination of the flavor of the $B_{tag}$, as described 
in detail in~\cite{BabarPub0008}.  
\par
Each event with a \CP\ candidate is assigned a $\Bz$ or $\Bzb$ tag if 
the rest of the event ({\it i.e.,} with the daughter
tracks of the $B_{CP}$ removed) satisfies the criteria from one of several 
tagging categories. 
In other words, a \Bz\ tag indicates that the $B_{CP}$ candidate was in a \Bzb\ state 
at $\deltat=0$; a \Bzb\ tag indicates that the $B_{CP}$ candidate was in a \Bz\ state.
\par
Three tagging categories rely on the presence of a fast lepton 
({\tt Lepton} categories) and/or one or more charged kaons in the event
({\tt Kaon} category).  
Two categories, called neural network categories ({\tt NT1} and {\tt NT2}), are based upon
the output value of a neural network algorithm applied to events 
that have not already been assigned to lepton or kaon tagging categories.
\item 
Measurement of the dilution factors ${\cal D}_i$ from the data  
for the different tagging categories, as described in detail
in~\cite{BabarPub0008}.
\par
The figure of merit for each tagging category is the effective tagging efficiency  
$Q_i = \varepsilon_i \, \left( 1 - 2\mistag_i \right)^2$, where $\varepsilon_i$ 
is the fraction of events assigned to category $i$ and 
$\mistag_i$ is the mistag fraction.
\par
The mistag fractions are measured directly in events 
in which one \Bz\ candidate, called the $B_{rec}$, is fully reconstructed 
in a flavor eigenstate mode.
The flavor-tagging algorithms are applied to the 
rest of the event, which constitutes the potential $B_{tag}$, assuming the $B_{rec}$ is 
properly tagged.
\begin{table}[!tb]
\caption{
Mistag fractions measured from 
a maximum-likelihood fit to the time distribution for the fully-reconstructed \Bz\ sample.
The {\tt Electron} and {\tt Muon}  categories are grouped into 
one {\tt Lepton} category.  
The uncertainties on $\varepsilon$ and $Q$ are statistical only.
} 
\vspace{0.3cm}
\begin{center}
\begin{tabular}{|l|c|c|c|} \hline
Tagging Category & $\varepsilon$ (\%) & $\mistag$ (\%) & $Q$ (\%)       \\ \hline \hline
{\tt Lepton}     & $11.2\pm0.5$ & $9.6\pm1.7\pm1.3$   &  $7.3\pm0.7$  \\
{\tt Kaon}       & $36.7\pm0.9$ & $19.7\pm1.3\pm1.1$  &  $13.5\pm1.2$  \\
{\tt NT1}        & $11.7\pm0.5$ & $16.7\pm2.2\pm2.0$  &  $5.2\pm0.7$  \\
{\tt NT2}        & $16.6\pm0.6$ & $33.1\pm2.1\pm2.1$  &  $1.9\pm0.5$  \\  \hline \hline
all              & $76.7\pm0.5$ &                     &  $27.9\pm1.6$ \\ 
\hline
\end{tabular}
\end{center}
\label{tab:TagMix:mistag}
\end{table}

The mistag fractions and the tagging efficiencies 
obtained by combining the results from maximum likelihood fits to the
time distributions in the \Bz\
hadronic and semileptonic samples are summarized 
in Table~\ref{tab:TagMix:mistag}. 
\item
Extraction of the amplitude of the \CP\ asymmetry and the value of \stwob\
with an unbinned maximum likelihood fit.  

\begin{figure}[t]
    \centering
 \begin{tabular}{lr}   
    \includegraphics[height=8cm,width=8cm]{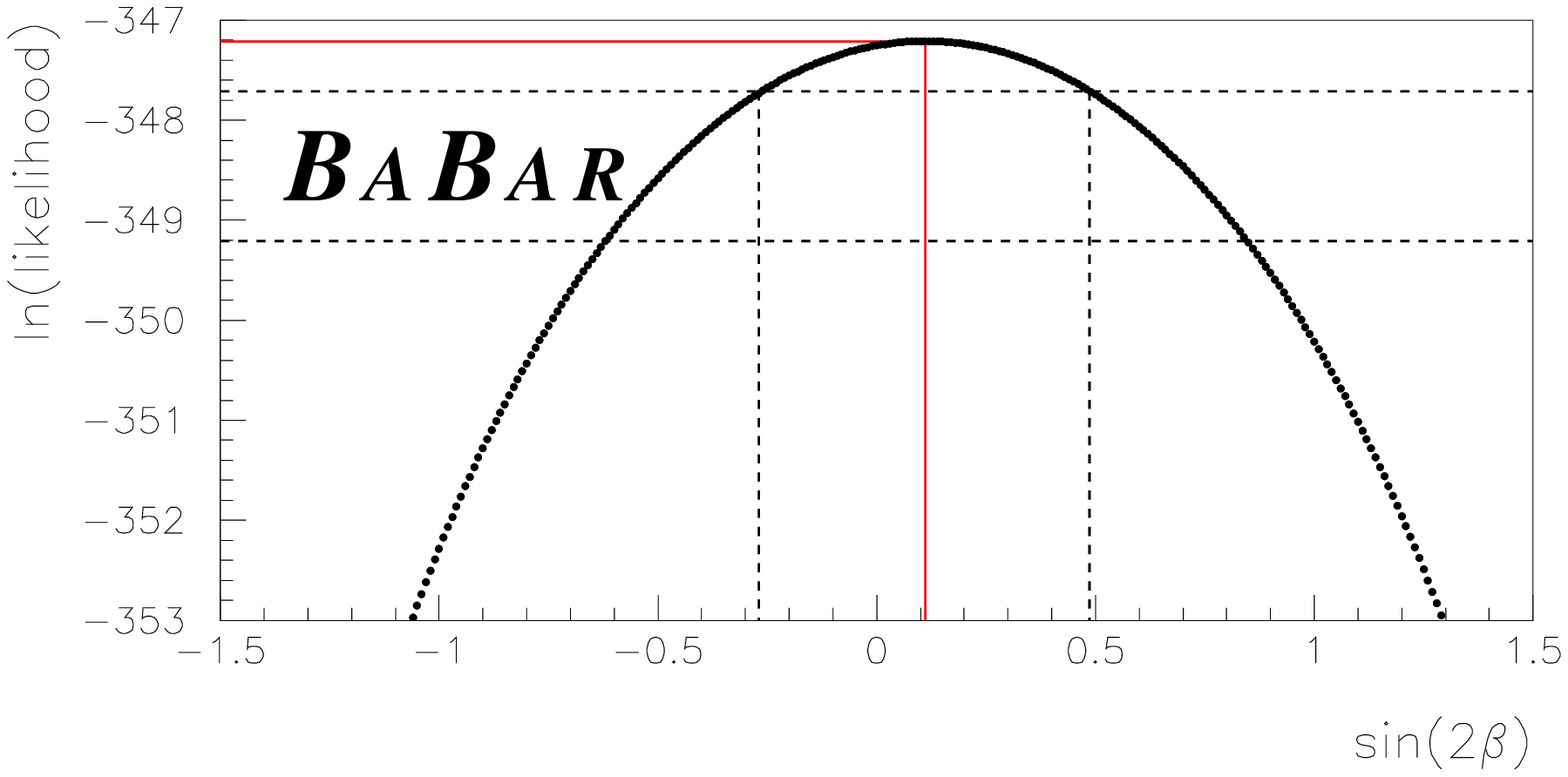}&
    \includegraphics[height=7cm,width=9cm]{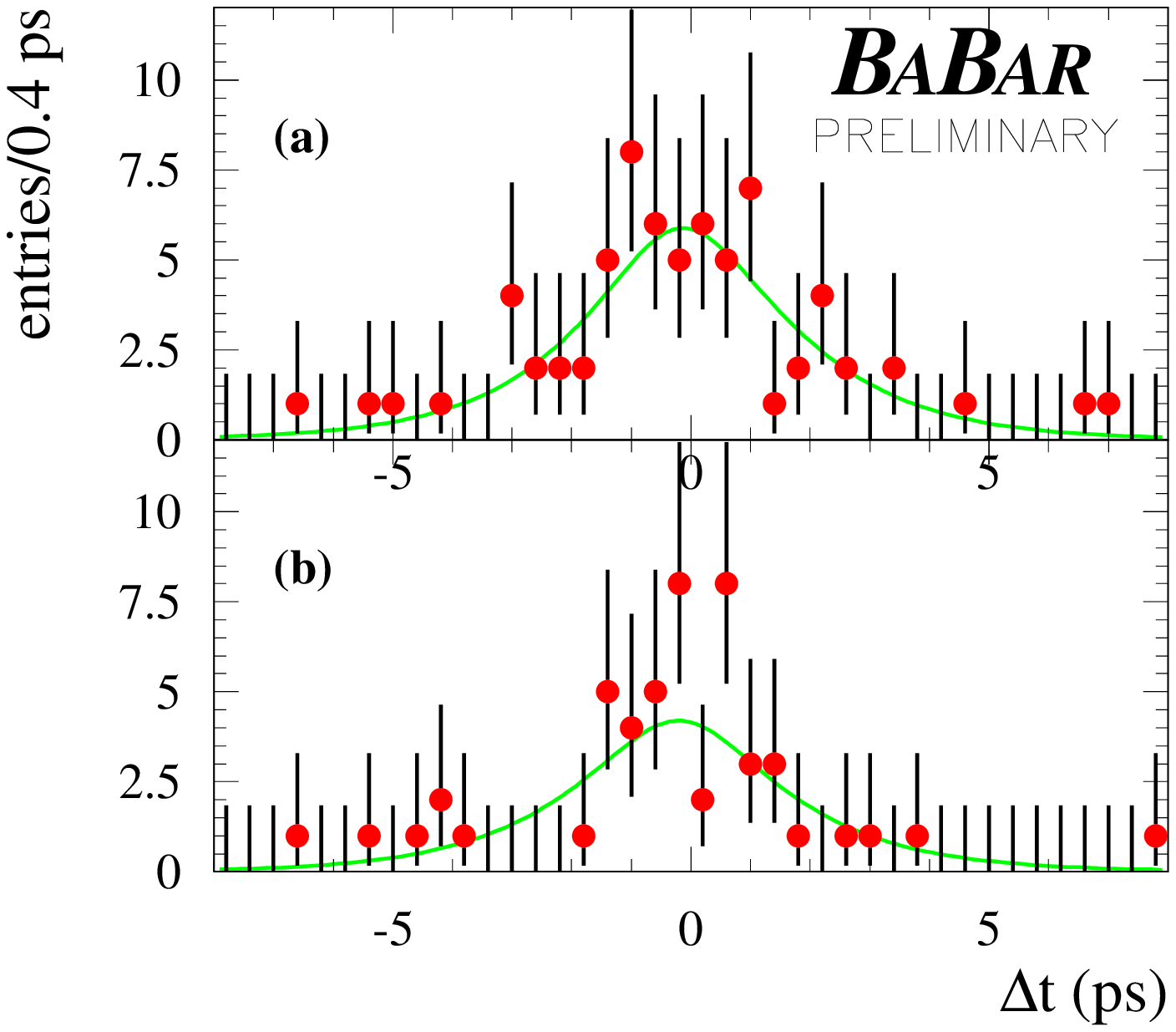}\\
\end{tabular}
 \caption{\it
      Variation	of the log likelihood as a function of \stwob\ (left). 
The two horizont al dashed lines indicate changes in the log likelihood 
corresponding to one and two statistical standard deviations.
Distribution of \deltat\ for (a) the \Bz\ tagged events and (b) the  \Bzb\ tagged events 
in the \CP\ sample (right).      \label{fig:likelihood} }
\end{figure}

The maximum-likelihood fit for \stwob, using the full tagged sample of $\Bz/\Bzb \to \jpsi \KS$ and $\Bz/\Bzb \to \psitwos \KS$ 
events, gives:
\begin{equation}
\result  .
\end{equation}

\begin{table}[!t]
\caption{
Summary of systematic uncertainties.  
The different contributions to the systematic 
error are added in quadrature.
} 
\vspace{0.3cm}
\begin{center}
\begin{tabular}{|l|c|} \hline
 Source of uncertainty    &  Uncertainty on \sb \\ \hline \hline
 
 uncertainty on $\tau_\Bz$                         &   0.002     \\
 uncertainty on \deltamd                           &   0.015     \\
 uncertainty on \deltaz\ resolution for \CP\ sample  &   0.019     \\ 
 uncertainty on time-resolution bias for \CP\ sample               &   0.047     \\ 
 uncertainty on measurement of mistag fractions        &  0.053    \\   
 different mistag fractions for \CP\ and non-\CP\ samples       &   0.050     \\
 different mistag fractions for \Bz\ and \Bzb\     &   0.005     \\
 background in \CP\ sample                         &   0.015     \\
\hline \hline
 total systematic error                            & {\bf 0.091 }   \\ 
\hline 
\end{tabular}
\end{center}
\label{tab:systematics}
\end{table}

For this result, the \Bz\ lifetime and \deltamd\ are fixed to the current best 
values~\cite{PDG2000}.
The log likelihood is shown as a function of \stwob\ 
and the \deltat\ distributions for \Bz\ and \Bzb\ tags are shown 
in Fig.~\ref{fig:likelihood}.
The contributions to the systematic uncertainty are summarized
in Table~\ref{tab:systematics}. They are added in quadrature to obtain the
total systematic error. 
\par
Improvements in the \sb\ measurement are expected in the near 
future with the accumulation and analysis of more data and further systematic studies.

\end{itemize}
\section{Measurements of charged and neutral \B\ meson lifetimes and $\Bz\Bzb$ oscillations}
\label{sec:mixing}
These measurements can be used to test theoretical models of heavy--quark decays 
and to constrain the Unitarity Triangle (via the sensitivity to the value of the CKM
matrix element $V_{td}$).
\par
One \B~($\B_{rec}$) is fully reconstructed in an all-hadronic
($\Bz \to D^{(*)-} \pi^+$, $D^{(*)-} \rho^+$,
$D^{(*)-} a_1^+$, $\jpsi \Kstarz$ and
$\Bu \to \overline{D}^{(*)0} \pip$,
$\jpsi K^+$, $\psitwos K^+$)
or semileptonic decay mode ($\bztodstarlnu$)~\footnote{Throughout this paper, 
conjugate modes are implied.}.  A total of about 2600 \Bz\
and a similar number of charged \B\ are
reconstructed in the hadronic modes, with a mean purity of $\sim 90\%$. 
The background is mainly combinatorial. About 7500 \Bz\
are reconstructed in the semileptonic modes, with a mean purity of $\sim
84\%$. Backgrounds to the semi-leptonic mode are due to combinatorics, 
wrong leptons, $c\bar c$ events, and charged \B\ decays from
$B^+\rightarrow D^{**0} l^+ \nu$. 
\par
The separation between the two \B\ vertices ($ z_{rec}$ and $z_{TAG}$
for the reconstructed and tagged vertex, respectively) along the boost direction,
$\Delta z = z_{rec} - z_{tag}$, is measured and used to
estimate the decay time difference, $\Delta t = \Delta z/\beta_z c$.
The $\Delta t$ resolution is dominated by the precision on the
$\B_{tag}$ vertex, and has little dependence on the decay mode of
the $\B_{rec}$.  
\subsection{Lifetime Measurements}
The \Bz\ and \Bu\ lifetimes are extracted from a simultaneous
unbinned maximum likelihood fit to the $\Delta t$ distributions of the
signal candidates, assuming a common resolution function.
Only hadronic modes have been used. An empirical
description of the $\Delta t$ background shape is assumed, using \mes\
sidebands with independent
parameters for neutral and charged mesons. Fig. \ref{fig:lifetime}
shows the $\Delta t$ distributions with the fit result superimposed.

\begin{figure}[t]
    \centering
 \begin{tabular}{lr}   
    \includegraphics[height=8cm,width=8cm]{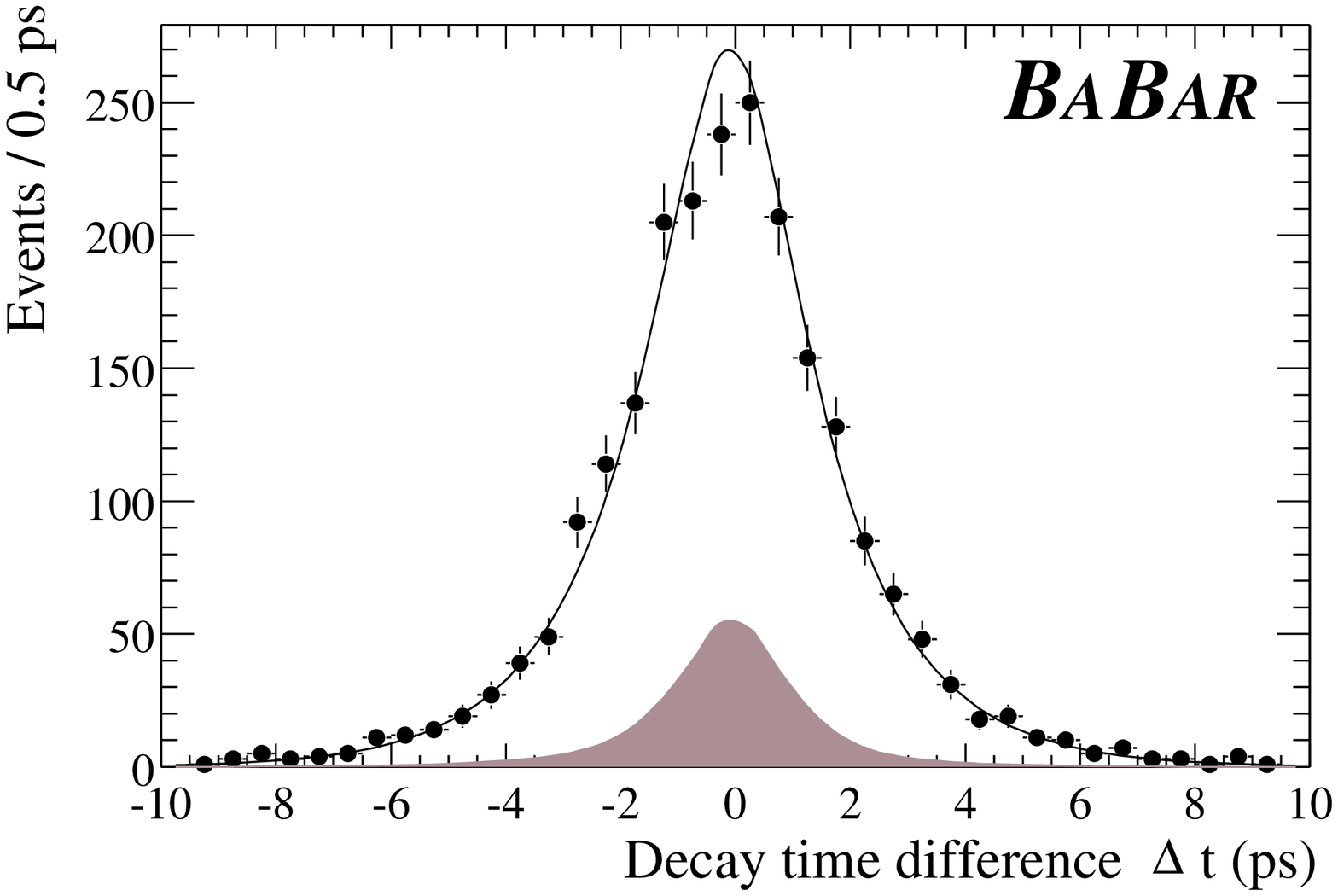}
    \includegraphics[height=8cm,width=8cm]{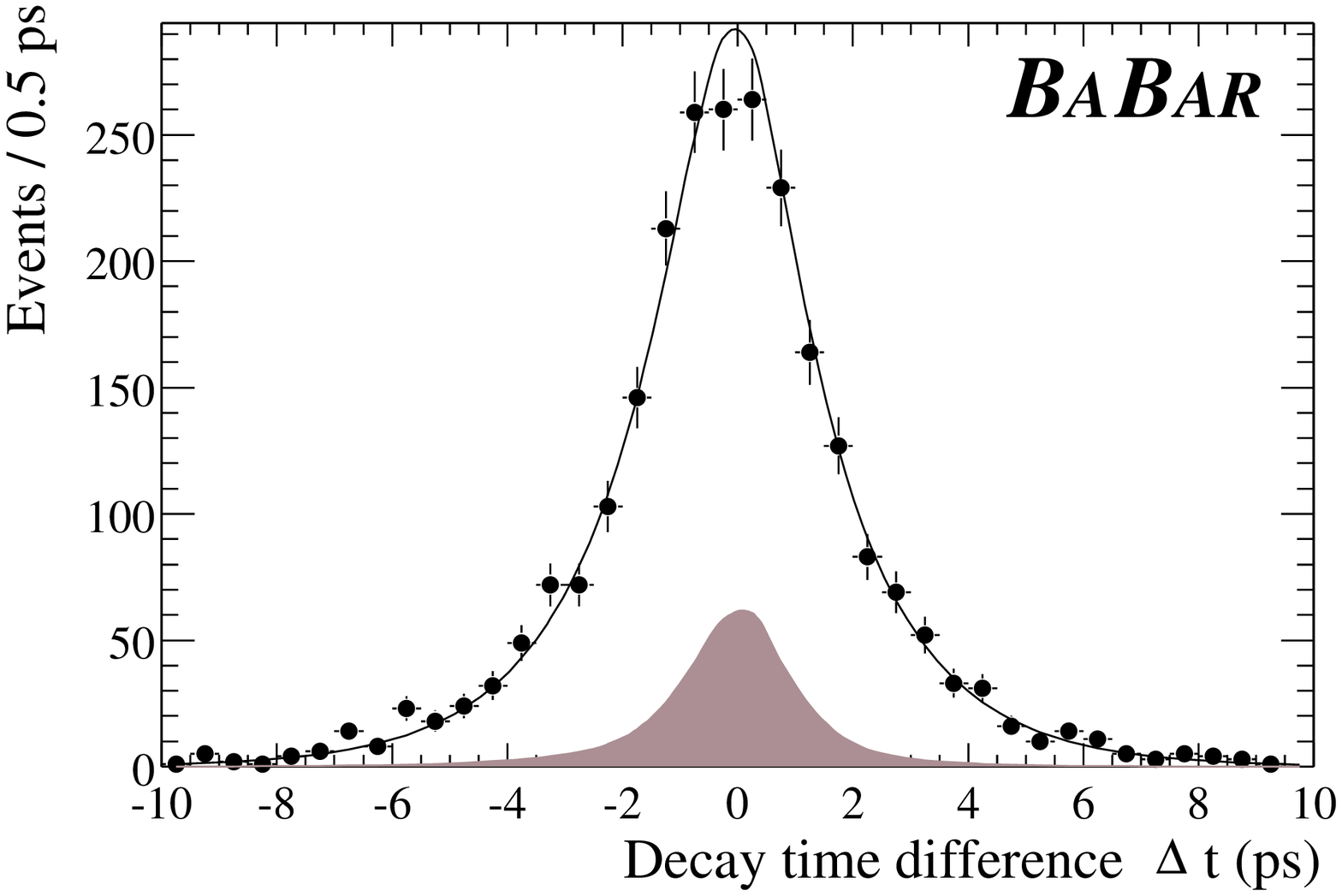}
\end{tabular}
\caption{\it $\Delta t$ distributions for \Bz/\Bzb\ (left) and \Bu/\Bub\
  (right) candidates in the signal region (\mes$>5.27$ \gevcc). The result of the lifetime
fit is superimposed. The background is shown by the hatched area.}
\label{fig:lifetime}
\end{figure}\medskip

\subsection{Time--dependent \BzBzb\ mixing}

A time-dependent \BzBzb\ mixing measurement 
requires the determination of the flavor of both \B\ mesons.
Considering the \BzBzb\ system as a whole, one can classify the tagged events 
as {\em mixed} or {\em unmixed} depending on whether the $B_{tag}$ is tagged 
with the same flavor as the $B_{rec}$ or with the opposite flavor. 
\par
From the time-dependent rate of mixed ($N_{mix}$) and unmixed
($N_{unmix}$) events, the mixing asymmetry 
$a(\Delta t) = (N_{unmix}-N_{mix})/(N_{unmix}+N_{mix})$ is
calculated as a function of $\Delta t$ and fit to the expected
cosine distribution.
A simultaneous unbinned 
likelihood fit to the mixing asymmetry frequency
and its amplitude allows the determination of both \deltamd\
and the mistag rates, $\mistag_i$, for the different tagging categories. 
The fit is performed simultaneously
in each tagging category, assuming a common resolution function. Fig. \ref{fig:mixing}
shows the $a(\Delta t)$ distributions with the fit result superimposed. 

\begin{figure}[t]
    \centering
 \begin{tabular}{lr}   
    \includegraphics[height=8cm,width=8cm]{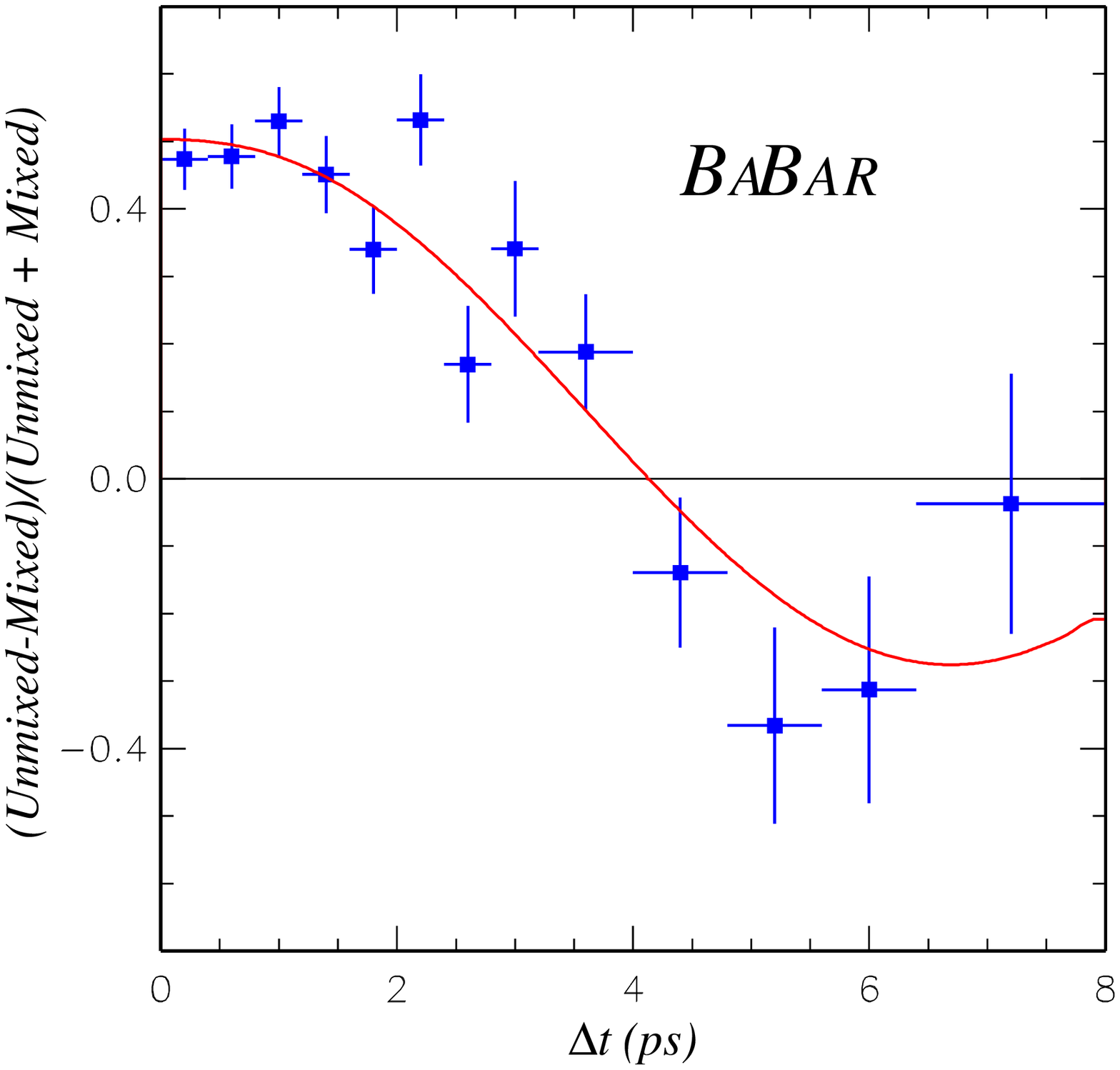}
    \includegraphics[height=8cm,width=8cm]{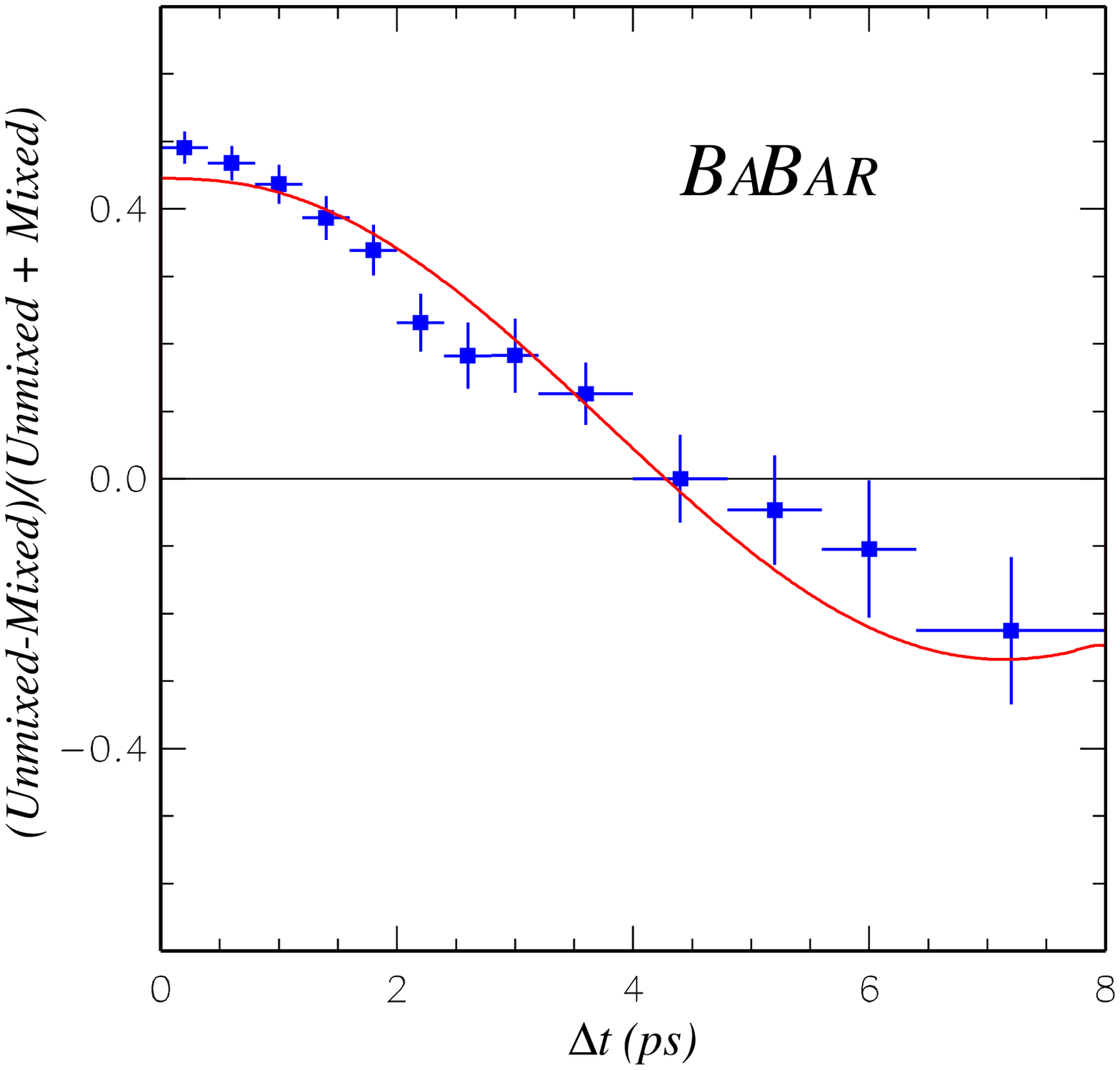}
\end{tabular}
\caption{\it Time-dependent asymmetry $a(\Delta t)$ between unmixed and
      mixed events for (left) hadronic $B$ candidates with $\mes >
      5.27$\gevcc and (right) for \btodstarlnu\ candidates.}
\label{fig:mixing}
\end{figure}\medskip

\subsection{Results}

The preliminary results for the \B\ meson lifetimes are
$\tau_{\Bz} = 1.506 \pm 0.052\ {\rm (stat)} \pm 0.029\ {\rm (syst)}\ \ps$ and
$\tau_{\Bu} = 1.602 \pm 0.049\ {\rm (stat)} \pm 0.035\ {\rm (syst)}\ \ps$
and for their ratio is
$\tau_{\Bu}/\tau_{\Bz} = 1.065 \pm 0.044\ {\rm (stat)}\ \pm 0.021\ {\rm (syst)}.$
\par
From the hadronic \Bz\ sample we measure the \BzBzb\ oscillation frequency:
$  \Delta m_d  =  0.516 \pm 0.031\ ({\rm stat})  \pm 0.018  ({\rm
  syst})\  \hbar {\rm ps}^{-1}$
and from the $D^{*-}\ell^+\nu$ sample the result is
$
  \Delta m_d   =   0.508 \pm 0.020\ ({\rm stat}) {}\pm 0.022 ({\rm
  syst})\ \hbar  {\rm ps}^{-1}$
Combining the two \deltamd\ results, we obtain the preliminary result: 
$\Delta m_d = 0.512 \pm  0.017 ({\rm stat})  \pm 0.022 ({\rm syst})\ \hbar {\rm ps}^{-1}.$
The mistag rates and $\Delta t$ resolution
function extracted from these fits to the data are used 
in the \babar\ \CP\ violation asymmetry analysis \cite{BabarPub0001}.
The above results are consistent with previous
measurements~\cite{PDG2000} and  are of
similar precision. The mixing result is compatible with 
a \babar\ measurement using di--leptons \cite{BabarPub0010}. 

\section{Conclusions}
We have presented \babar's first measurement of the \CP-violating 
asymmetry parameter \stwob\ in the $B$ meson system:
\begin{equation}
\result .
\end{equation}  
Our measurement is consistent with the world average $\stwob = 0.9\pm0.4$~\cite{PDG2000}, 
and is currently limited by the size of the \CP\ sample.
\par We have presented also the time--dependent mixing and lifetime measurements, 
performed for the first time at the \FourS.
\par Other important competitive measurements have been achieved by \babar.
\par
All measurement are foreseen to be improved in the near future, 
with the accumulation and study of more data and further systematic studies.

\end{document}